# The Stokes-Einstein-Debye relation in ortho-terphenyl liquid

Gan Ren (任淦)[†]

School of Science, Civil Aviation Flight University of China, Guanghan 618307, China

**Abstract**

The Stokes-Einstein-Debye (SED) relation is proposed to be breakdown in supercooled liquids by many studies. However, the conclusions are usually drawn by testing some variants of the SED relation rather than its original form. In this work, the rationality of the SED relation and its variants is examined by performing molecular dynamics simulations with the Lewis-Wahnstrom model of ortho-terphenyl (OTP). The results indicate the original SED relation is valid for OTP but the three variants are all breakdown. The inconsistency between SED relation and its variants is attributed to the adopted assumptions and approximations, heterogeneous dynamics and the interactions among molecules. Therefore, care should be taken when employing its variants to judge the validity of the SED relation in supercooled liquids.

**Keywords:** Stokes-Einstein-Debye relation, Stokes-Einstein relation, supercooled liquids, Stokes' formula

## 1. Introduction

The Stokes-Einstein-Debye (SED) relation [1] $D_r = k_B T/\varsigma$ correlates the rotational diffusion constant $D_r$, rotational friction coefficient $\varsigma$, Boltzmann constant $k_B$ and temperature $T$. The $\varsigma$ can be described by the Stokes' formula $\varsigma = 8\pi\eta a^3$ for a rigid sphere with radius $a$ moving in a fluid with viscosity $\eta$. So the SED relation can be expressed by $D_r = k_B T/8\pi\eta a^3$.

[†]E-mail address: rengan@alumni.itp.ac.cn

Debye [1, 2] proposed the $\varsigma$ is correlated with rotational relaxation time $\tau_r$, where $\tau_r$ is determined by the decay of the *n*-th degree Legendre polynomials. Then the SED can be expressed as the variant $D_r = 1/[\tau_{rn} n(n+1)]$. If one assume the effective hydrodynamic radius $a$ for soft particle to be also a constant like a rigid particle, one gets another variant of the SED as $D_r \sim T/\eta$ [3], where "~" means proportional. Furthermore one can get the variant $D_r \sim T/\tau_t$ based on the approximate relation $\eta = G_\infty \tau$ [4], where $G_\infty$ is the instantaneous shear modulus presumed to be a constant, and $\tau_t$ is the structural relaxation time determined by the decay of the self-intermediate scattering function. Therefore, the SED has at least four forms, the original form $D_r = k_B T / 8\pi\eta a^3$ and the three variants $D_r = 1/[\tau_{rn} n(n+1)]$, $D_r \sim T/\eta$ and $D_r \sim T/\tau_t$. Moreover, if $D_r = 1/[\tau_{rn} n(n+1)]$ is satisfied, $m(m+1)\tau_{rm} = n(n+1)\tau_{rn}$ should be established for different *n*, *m*. On the other hand, if the Stokes-Einstein (SE) relation $D_t = k_B T / 6\pi\eta a$ and the assumption of constant $a$ are satisfied, one can show the SED variant can be expressed as the ratio $D_t \tau_{rn}$, $D_t/D_r$ and $\tau_{rn} T/\eta$, etc are constant, where $D_t$ is the translational diffusion constant.,

By testing above SED variant, many studies proposed the SED relation is invalid in supercooled liquids due to the heterogeneous dynamics under supercooled state. A deviation from $\tau_{r1}/\tau_{r3} = 3$ is observed in both analytical models and experiments [5-7]. De Michele and Leporini [8] found the ratio $n(n+1)\tau_{rn}/2\tau_{r1} \neq 1$ for $n = 2, 3, 4$ in a supercooled liquid of rigid dumbbell model interacting via a Lennard-Jones potential. The $n(n+1)\tau_{rn}/2\tau_{r1}$ is found to be firstly decreased and then increased with cooling; similar changes are also observed in $n(n+1)D_r \tau_{rn}$. The dumbbell model interacting via a repulsive ramp like potential [9] shows that the $\tau_{r1} D_t$ is not a constant but temperature and density dependent. Similar phenomena are also observed in the simulations of the supercooled SCP/E water [10-12].

The ratio $D_r \tau_t / T$ is almost a constant in supercooled SPC/E water [4] with temperature range 280-350K but increases with cooling below 280K. However, a fractional form $D_r \sim (T/\tau_t)^\xi$ with $\xi = 0.75$ is observed for the whole simulated temperature range. The ratio $D_t/D_r$ is not a constant but decreases with decreasing temperature. For TIP4P water model, Kawasaki and Kim [3] found $D_t/D_r$ is also decreased with cooling but $D_r \tau_{rn}$ with $n = 1, 2, 3, 6$ show the reverse trend; $\tau_{r6} T/\eta$ is almost established but a fractional form $\tau_1 \sim (\eta/T)^\xi$ is observed with $\xi = 0.8$. The scaled ratio $D_t/D_r$ is almost equal to 1.0 in OTP [13] within 260-346K but deviates from 1.0 with cooling. Moreover, a controversial result is observed in OTP that the simulated $D_t/D_r$ shows an opposite trend with the data deduced from $\tau_{r2}$. The breakdown of variant by combining of the SED and SE relation is usually attributed to the decoupling of the transitional and rotational motion.

Although there exist so many studies suggest the breakdown of SED relation in supercooled liquids, no study directly tests the original SED relation $D_r = k_B T / 8\pi \eta a^3$. This is questionable because the equivalence of the variants to the original SED form is on the basis of various assumptions, while there are some evidences showing that those assumptions may not be valid all the time. For instance, the $a$ for organic molecules varies with volume fraction in their diluted solutions [14], and the behavior of ions in aqueous solutions is observed to deviate from the SE relation by taking $a$ as a constant but the original SE relation actually holds if $a$ is allowed to change [15-19]. Moreover, there exist simulations [20, 21] indicate the Einstein relation $D_t = k_B T/\alpha$ is valid for several supercooled liquids within certain temperature range, and the $a$ should be varied with temperatures on the basis of the validity of the Stokes' formula $\alpha = 6\pi \eta a$. And the relation $\eta = G_\infty \tau_t$ is approximately established only when the memory effect is exponential [2]; however, the structural relaxation follows non-exponential decay in supercooled liquids for the dynamic heterogeneity [22-24]. For the variants given by combination of the SED and SE relation variants, there exist many studies show the breakdown of SE relation variants [4, 25-29]. So the validity of the SED relation is still elusive and one should consider it from the original form. In this work, we explore the SED relation from its original form and variants to verify its validity by performing molecular dynamics (MD) simulations with the Lewis-Wahnstrom model of ortho-terphenyl (OTP) [28, 30].

## 2. Simulation details and analysis methods

The present work is based on our previous work [21], the adopted OTP model [28, 30] and simulation details are the same. The frictional coefficient $\alpha$ and viscosity $\eta$ are directly taken from ref. [21], which are also plotted in Fig. 1(a) and Fig. 1(b) for convenience. The three sites of OTP molecule are named $A$, $B$ and $A$, respectively. Their charges are $q_A = 0, q_B = 0$. To explore the possible influences of torque introduced by interaction among molecules on the SED variants, other two systems are simulated with charges $q_A = 0.02, q_B = -0.04$ and $q_A = 0.04, q_B = -0.08$ in unit $e$, respectively. To improve statistics, seven independent trajectories have been simulated to determine the structural relaxation time $\tau_t$, rotational diffusion constant $D_r$, rotational correlation time $\tau_{rn}$ and rotational non-Gaussian parameter $\alpha_2(t)$.

The $\tau_t$ is described by the self-intermediate scattering function [31]

$$F_s(k,t) = \frac{1}{N} \sum_{j=1}^{N} \left\langle e^{i\vec{k}\cdot[\vec{r}_j(t)-\vec{r}_j(0)]} \right\rangle \tag{1}$$

where $N$ is the number of molecules, wavevector $k = 14.5 nm^{-1}$ corresponding to the first maximum of the static structure factor, $\vec{r}_j(t)$ is the position of center of mass for the $j$-th molecule, $\langle\rangle$ denotes time average, and $\tau_t$ is determined by $F_s(k,\tau_t) = e^{-1}$.

The rotational diffusion constant $D_r$ is calculated via its asymptotic relation with the rotational mean square displacement (RMSD) [3, 4]

$$D_r = \lim_{\Delta t \to \infty} \frac{\left\langle \vec{\varphi}^2(\Delta t) \right\rangle}{4\Delta t} \tag{2}$$

where $\left\langle \vec{\varphi}^2(\Delta t)\right\rangle = 1/N\sum_{i=1}^{N}\left|\vec{\varphi}_i(t+\Delta t)-\vec{\varphi}_i(t)\right|^2$ is the RMSD for the displacement $\vec{\varphi}_i(\Delta t)$. The $\vec{\varphi}_i(\Delta t) = \int_{t}^{t+\Delta t}\vec{\omega}_i(t)dt$, where the angular velocity $\vec{\omega}_i(t)$ of $i$-th molecule is in magnitude $\left|\vec{\omega}_i(t)\right| = \cos^{-1}\left[\vec{e}_i(t)\cdot\vec{e}_i(t+\Delta t)\right]$ and direction $\vec{e}_{\omega_i} = \vec{e}_i(t)\times\vec{e}_i(t+\Delta t)/\Delta t$, $\vec{e}_{\omega_i}$ is the unit vector of the angular bisector of $i$-th OTP molecule. A time interval $\Delta t = 0.01 ps$ is adopted to calculate the RMSD.

The rotational correlation time $\tau_{rn}$ is calculated via the rotational correlation function [1, 3]

$$C_n(t) = \frac{1}{N}\sum_{i=1}^{N}\left\langle P_n\left[\vec{e}_i(t)\cdot\vec{e}_i(0)\right]\right\rangle \quad (3)$$

where $P_n(x)$ is the $n$-th order Legendre polynomial, and $\tau_{rn}$ is determined by $C_n(\tau_{rn}) = e^{-1}$.

The rotational dynamics are heterogeneous and are characterized by the rotational non-Gaussian parameter [32]

$$\alpha_2(t) = \frac{3\left\langle \vec{\varphi}^4(t)\right\rangle}{5\left\langle \vec{\varphi}^2(t)\right\rangle^2} - 1 \quad (4)$$

## 3. Results and discussion

To examine the SED relation and its variants, the viscosity $\eta$, frictional coefficient $\alpha$, rotational relaxation time $\tau_{rn}$ for $n =$ 1, 2, 6, structural relaxation time $\tau_t$ and rotational diffusion constant $D_r$ at different temperature $T$ are calculated and plotted in Fig. 1.

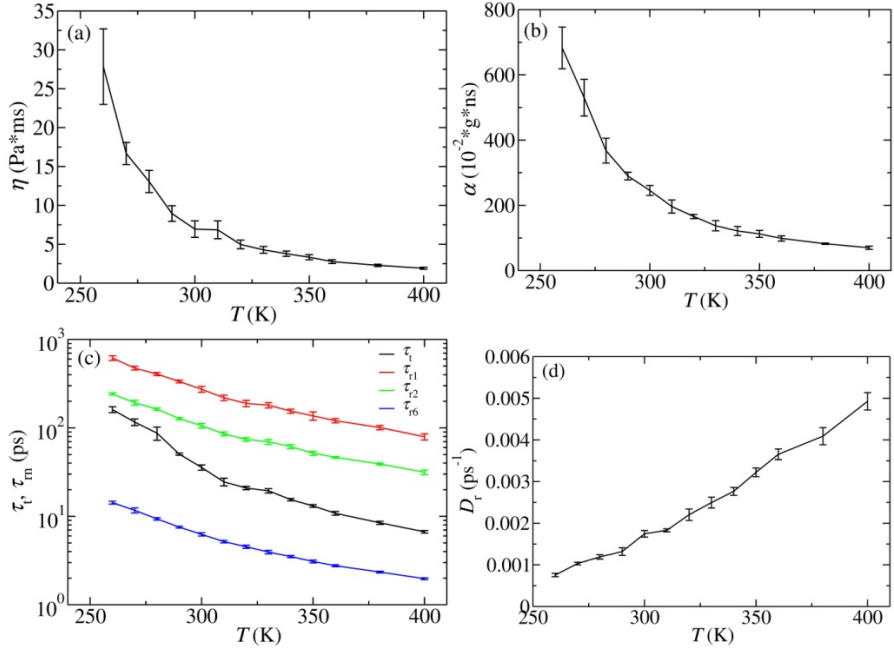

**Figure 1.** The $\eta$, $\alpha$, $\tau_{rn}$ for $n = 1, 2, 6$, $\tau_t$ and $D_r$ as a function of $T$: (a) $\eta$ vs $T$; (b) $\alpha$ vs $T$; (c) $\tau_{rn}$, $\tau_t$ vs $T$; (d) $D_r$ vs $T$.

The variant $D_r \sim T/\eta$ behaves as a fractional form $D_r \sim (T/\eta)^{\xi_1}$ with $\xi_1 = 0.61$ as shown in Fig. 2(a), which deviates 0.39 from the exact $\xi_1 = 1.0$ and implies the breakdown of $D_r \sim T/\eta$. The exponent $\xi_1 = 0.61$ is smaller than the $\chi \approx 0.9$ in $D_t \sim (T/\eta)^{\chi}$ in ref. [21], which implies $D_t/D_r \sim a^2 \sim (T/\eta)^{\chi - \xi_1}$ should decrease with decreasing temperature. The decreasing of $D_t/D_r$ with cooling is observed in the simulation of the TIP4P water [3] and OTP [13]. Similar breakdown is observed in $D_r \sim T/\tau_t$ but the fractional form $D_r \sim (T/\tau_t)^{\xi_2}$ with an exponent $\xi_2 = 0.49$ as plotted in Fig. 2(b). The $\xi_1 = 0.61$ and $\xi_2 = 0.49$ show the relation $\eta = G_\infty \tau$ is not exact, the similar are observed in testing SE relation [17, 20, 28].

The $D_r = 1/[\tau_{rn} n(n+1)]$ is tested by $D_r \sim \tau_{rn}^{-1}$ for $n = 1, 2, 6$. The three are all in fractional forms as $D_{rn} \sim \tau_{rn}^{-\xi_3}$, and the three exponents $\xi_3 \approx 0.9$ as plotted in Fig. 2(c). The breakdown is small and $D_r \sim \tau_{rn}^{-1}$ is approximately valid. The result is different from the TIP4P water [3], one observes that $D_r \sim \tau_{r1}^{-\xi_3}$ is invalid with $\xi_3 \approx 0.8$ and $D_r \sim \tau_{r6}^{-\xi_3}$ is almost established with $\xi_3 \approx 1.0$. Comparing $D_t \sim (T/\eta)^\chi$ and $D_r \sim \tau_{rn}^{-\xi_3}$ for $n = 2$, the two exponents are almost equal. However, the increases of $\eta$ are much faster than the increases of $\tau_{r2}$ with cooling as shown in Fig. 1, and one can explain the observed opposite trend for $D_t/D_r$ vs $T$ in the data deduced from $\tau_{r2}$ and the simulated [13]. We assume the Stokes' formula $\alpha = 6\pi\eta a$ is established, and deduce the $a$ by $a \sim \alpha/\eta$, then the original form $D_r = k_B T/8\pi\eta a^3$ is tested by $D_r \alpha^3 \sim (T\eta^2)^{\xi_4}$. The directly fitted exponent $\xi_4$ is 1.06, which is so close to the exact result marked by the red dashed with $\xi_4 = 1.0$ as plotted in Fig. 2(d). It indicates $D_r = k_B T/8\pi\eta a^3$ is valid. And we can see the breakdown of $D_r \sim T/\eta$ is attributed to the adopted assumption of constant $a$ is not valid.

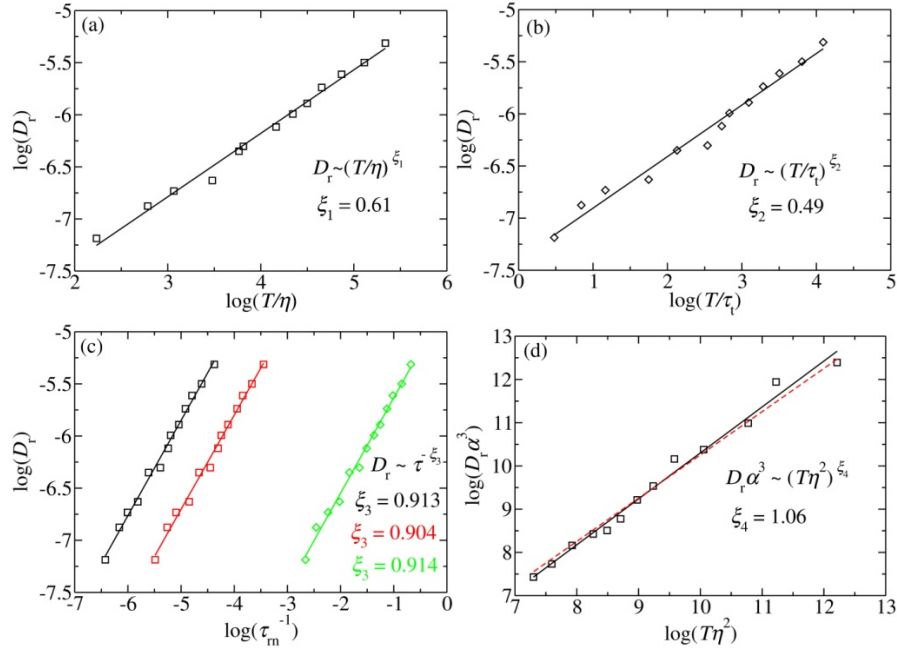

**Figure 2.** Verification of the SED relation and its variants: (a) $D_r \sim T/\eta$; (b) $D_r \sim T/\tau_t$; (c) $D_r \sim \tau_{rn}^{-1}$, black, red and green symbol are for $n = 1, 2, 6$, respectively; (d) $D_r \alpha^3 \sim T\eta^2$. The calculated data are represented by symbols and the solid lines are the fitting. The red dashed line in (d) is fitted with $\xi_4 = 1.0$.

The above results indicate the $D_r \sim T/\eta$ and $D_r \sim T/\tau_t$ are definitely invalid, and whether $D_r \sim \tau_{rn}^{-1}$ is really breakdown. If $\vec{\varphi}_i(\Delta t)$ follows Gaussian distribution, the $D_r \sim \tau_{r1}^{-1}$ is an exact result due to $\langle P_1(\cos\varphi)\rangle = \langle e^{i\varphi}\rangle = e^{-\langle\varphi^2\rangle/2}$. The rotational dynamics is also observed to be heterogeneous as the translational dynamics [8, 32-34]. The non-Gaussian parameter $\alpha_2(t)$ plotted in Fig. 3(a) is similar as the observed in SPC/E water [32], which deviates from zero and the maximum increases with cooling. The system has more heterogeneous dynamics at a lower temperature. So the $D_r \sim \tau_{r1}^{-1}$ should be invalid.

On the other hand, if molecule rotates without external torque, the probability distribution of the chosen unit vector $\vec{e}$ is $\rho[\vec{e}(t)] = \sum_{n,m} e^{-n(n+1)D_r t} Y_n^m[\vec{e}(t)] Y_n^{-m}[\vec{e}(0)]$ [2], and $D_r = 1/[\tau_{rn} n(n+1)]$ is an exact result, where $Y_n^m$ is the spherical harmonic function. Because of the dynamic heterogeneity, the system is heterogeneous both in dynamics and structure, and the more mobile particles are likely to from a string structure [24, 32]. So the net torque applied on a molecule may not be zero due to the interaction among molecules even without external torque. The most obvious torque is the molecular dipole couples to an electric field [1].

To explore the possible influence introduced by the interaction among molecules, we simulate two other systems consisted of polar molecules to observe the differences with a larger interaction. In order not to have a significant effect on the system, the three sites of OTP in the two systems are taken small electric charges with $q_A = 0.02, q_B = -0.04$ and $q_A = 0.04, q_B = -0.08$, respectively. Fig. 3(b) shows the $\alpha_2(t)$ for $q_A = 0.02, q_B = -0.04$ and $q_A = 0.04, q_B = -0.08$ have no much differences with the $q = 0$ system. However, the fractional form $D_{rn} \sim \tau_{rn}^{-\xi_3}$ for n = 1, 2, 6 has a smaller exponent $\xi_3$ for a larger $q$ as identified by the Fig. 3(c) and Fig. 3(d). For instance, the $\xi_3$ for n = 1 is decreased from 0.913 for $q = 0$ to 0.82 for $q_A = 0.02, q_B = -0.04$, and then is decreased to 0.78 for $q_A = 0.04, q_B = -0.08$. The results indicate the interaction among molecules may play a more important role in the breakdown of $D_r \sim \tau_{rm}^{-1}$ than the heterogeneous dynamics.

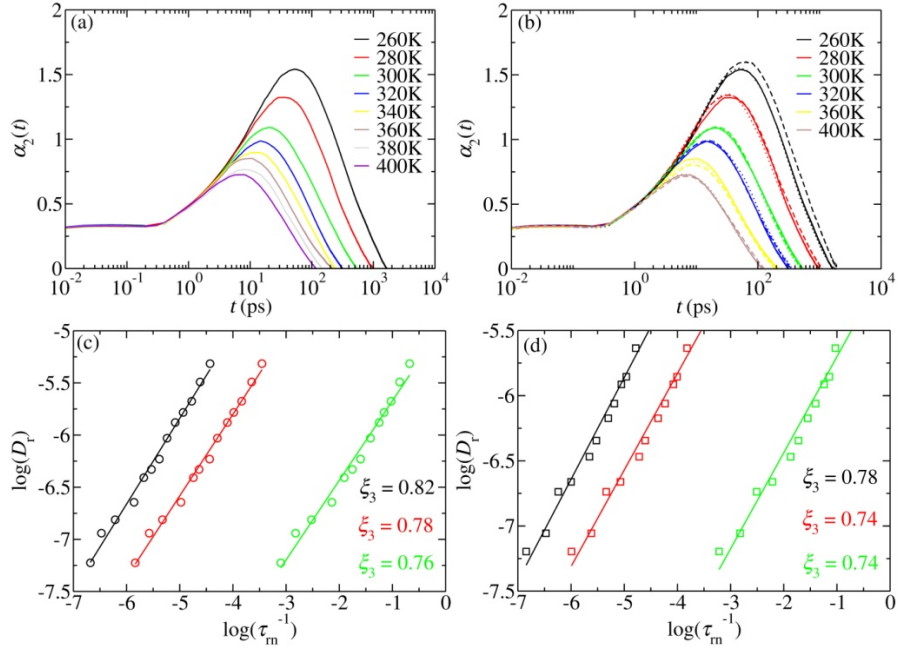

**Figure 3.** (a) non-Gaussian parameter $\alpha_2(t)$ for $q = 0$; (b) a comparison of non-Gaussian parameter $\alpha_2(t)$ for original and charged OTP, the solid line is for $q = 0$, the dotted is for $q_A = 0.02, q_B = -0.04$ and the dashed is for $q_A = 0.04, q_B = -0.08$. Testing the SED relation $D_r = 1/[\tau_{rn} n(n+1)]$ by $D_r \sim \tau_{rm}^{-\xi_3}$ for charged systems: (c) $q_A = 0.02, q_B = -0.04$; (d) $q_A = 0.04, q_B = -0.08$. The black, red and green symbol are for $n$ = 1, 2, 6, respectively. The solid line is the fitting.

To further verify the validity of $D_r \sim \tau_{rm}^{-1}$, the relation $n(n+1)\tau_{rn} = 2\gamma_n \tau_{r1}$ are adopted. If $D_r \sim \tau_{rm}^{-1}$ is valid, $\gamma_n = 1.0$ and otherwise $\gamma_n \neq 1.0$. Fig. 4 shows the $\gamma_2$ is increased from 1.194 for $q = 0$ to 1.288 for $q_A = 0.02, q_B = -0.04$, and then is increased to 1.295 for $q_A = 0.04, q_B = -0.08$. The $\gamma_6$ is increased from 0.495 for $q = 0$ to 0.584 for $q_A = 0.02, q_B = -0.04$, and then is decreased to 0.561 for $q_A = 0.04, q_B = -0.08$; the decrease may be attributed to the data error. Both $\gamma_2$ and $\gamma_6$ deviate significantly from $\gamma_n = 1.0$ and imply the breakdown of $D_r \sim \tau_{rm}^{-1}$. Moreover, the systems with $q \neq 0$ have a larger $\gamma_2$ and $\gamma_6$ than $q = 0$, and the larger $q$ system has a larger $\gamma_2$, which also signifies the importance of interaction for the breakdown of $D_r \sim \tau_{rn}^{-1}$. The result is consistent with the data plotted in Fig. 3. Combining the results given by Fig. 2, Fig. 3 and Fig. 4, we conclude the variant $D_r \sim \tau_{rn}^{-1}$ is indeed breakdown.

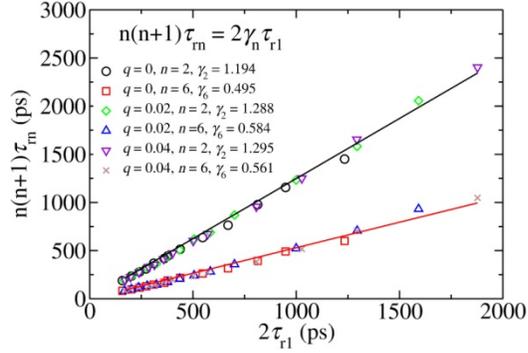

**Figure 4.** Verification of the validity of the SED relation $D_r = 1/[\tau_{rn} n(n+1)]$ by $n(n+1)\tau_{rn} = 2\gamma_n \tau_{r1}$ for different $q$.

## 4. Conclusions

In summary, we have examined the rationality of the SED relation and its variants in OTP liquids by performing MD simulations. Our results indicate $D_r \sim T/\eta$, $D_r \sim T/\tau_t$ and $D_r \sim \tau_{rn}^{-1}$ with $n = 1$, 2, 6 are all breakdown and in fractional forms. The breakdown of the variant $D_r \sim T/\eta$ is due to the assumption of constant $a$ is not satisfied. The breakdown of the variant $D_r \sim T/\tau_t$ is because it is based on $D_r \sim T/\eta$ and further adopt an approximation relation $\eta = G_\infty \tau$. The breakdown of $D_r \sim \tau_{r1}^{-1}$ is attributed to the rotational heterogeneous dynamics and displacements of angle deviations from Gaussian distribution. The deviations from $D_r \sim \tau_{rn}^{-1}$ with $n = 1$, 2, 6 get larger with a lager $q$. However, there exists no much difference in non-Gaussian parameter for the $q = 0$ and $q \neq 0$ system. So the interaction among molecules may play a more important role for the breakdown of $D_r \sim \tau_{rn}^{-1}$ than the deviation from Gaussian. Although the three variants are all breakdown, the original SED relation $D_r = k_B T/8\pi \eta a^3$ is valid after taking the changes of $a$ into account. The result is consistent with our previous work for the SE relation in OTP [21]. Moreover, no decoupling of translation and rotational motion is observed for OTP within 260-400K. Our simulations indicate that the $a$ is such an important parameter that is closely connected with conclusion drawn on the validity of SED relation, so one should be carefully evaluated the assumption of constant $a$ when testing the SED relation and care should be taken when using variants to test the SED relation.

## Acknowledgments

This work was supported by the National Natural Science Foundation of China (No. 12104502) and the Natural Science Foundation of Sichuan Province (No. 2023YFG0308). The author thanks professor Yanting Wang (Institute of Theoretical Physics, Chinese Academy of Science) for suggestions and supporting of computation resource on the Tian-2 supercomputer.

## References

[1] Debye P 1929 *Polar Molecules* (New York:The Chemical Catalog Company))
[2] Barrat J and Hansen J 2003 *Basic concepts for simple and complex liquids* (Cambridge: Cambridge University Press)
[3] Kawasaki T and Kim K 2019 Spurious violation of the Stokes–Einstein–Debye relation in supercooled water *Scientific Reports* 9 8118
[4] Mazza M G, Giovambattista N, Stanley H E and Starr F W 2007 Connection of translational and rotational dynamical heterogeneities with the breakdown of the Stokes-Einstein and Stokes-Einstein-Debye relations in water *Phys. Rev. E* 76 031203
[5] Kivelson D and Miles D 1988 Bimodal angular hopping model for molecular rotations in liquids *J. Chem. Phys.* 88 1925
[6] Kivelson D and Kivelson S A 1989 Models of rotational relaxation above the glass transition *J. Chem. Phys.* 90 4464
[7] Diezemann G, Sillescu H, Hinze G and Böhmer R 1998 Rotational correlation functions and apparently enhanced translational diffusion in a free-energy landscape model for the \ensuremath{\alpha} relaxation in glass-forming liquids *Phys. Rev. E* 57 4398
[8] De Michele C and Leporini D 2001 Viscous flow and jump dynamics in molecular supercooled liquids. II. Rotations *Phys. Rev. E* 63 036702
[9] Netz P A, Buldyrev S V, Barbosa M C and Stanley H E 2006 Thermodynamic and dynamic anomalies for dumbbell molecules interacting with a repulsive ramplike potential *Phys. Rev. E* 73 061504
[10] Stanley H E, Barbosa M C, Mossa S, Netz P A, Sciortino F, Starr F W and Yamada M 2002 Statistical physics and liquid water at negative pressures *Physica A: Statistical Mechanics and its Applications* 315 281
[11] Netz P A, Starr F W, Barbosa M C and Stanley H E 2002 Relation between structural and dynamical anomalies in supercooled water *Physica A: Statistical Mechanics and its Applications* 314 470
[12] Netz P A, Starr F, Barbosa M C and Stanley H E 2002 Translational and rotational diffusion in stretched water *J. Mol. Liq.* 101 159
[13] Lombardo T G, Debenedetti P G and Stillinger F H 2006 Computational probes of molecular motion in the

Lewis-Wahnström model for ortho-terphenyl *J. Chem. Phys.* 125

[14] Schultz S G and Solomon A K 1961 Determination of the Effective Hydrodynamic Radii of Small Molecules by Viscometry *The Journal of General Physiology* 44 1189

[15] Lee S H and Rasaiah J C 1994 Molecular dynamics simulation of ionic mobility. I. Alkali metal cations in water at 25° C *J. Chem. Phys.* 101 6964

[16] Lee S H and Rasaiah J C 1996 Molecular dynamics simulation of ion mobility. 2. Alkali metal and halide ions using the SPC/E model for water at 25 C *The Journal of Physical Chemistry* 100 1420

[17] Ren G, Chen L and Wang Y 2018 Dynamic heterogeneity in aqueous ionic solutions *Phys. Chem. Chem. Phys.* 20 21313

[18] Hubbard J and Onsager L 1977 Dielectric dispersion and dielectric friction in electrolyte solutions. I *J. Chem. Phys.* 67 4850

[19] Varela L M, García M and Mosquera V c 2003 Exact mean-field theory of ionic solutions: non-Debye screening *Phys. Rep.* 382 1

[20] Ren G and Wang Y 2021 Conservation of the Stokes–Einstein relation in supercooled water *Phys. Chem. Chem. Phys.* 23 24541

[21] Ren G 2022 The effective hydrodynamic radius in the Stokes–Einstein relation is not a constant *Commun. Theor. Phys.* 74 095603

[22] Berthier L 2011 Dynamic Heterogeneity in Amorphous Materials *Physics* 4

[23] Berthier L and Biroli G 2011 Theoretical perspective on the glass transition and amorphous materials *Rev. Mod. Phys.* 83 587

[24] Kob W, Donati C, Plimpton S J, Poole P H and Glotzer S C 1997 Dynamical Heterogeneities in a Supercooled Lennard-Jones Liquid *Phys. Rev. Lett.* 79 2827

[25] Kawasaki T and Kim K 2017 Identifying time scales for violation/preservation of Stokes-Einstein relation in supercooled water *Sci. Adv.* 3 e1700399

[26] Dehaoui A, Issenmann B and Caupin F 2015 Viscosity of deeply supercooled water and its coupling to molecular diffusion *Proc. Natl. Acad. Sci. U. S. A.* 112 12020

[27] Xu L, Mallamace F, Yan Z, Starr F W, Buldyrev S V and Eugene Stanley H 2009 Appearance of a fractional Stokes-Einstein relation in water and a structural interpretation of its onset *Nat. Phys.* 5 565

[28] Shi Z, Debenedetti P G and Stillinger F H 2013 Relaxation processes in liquids: Variations on a theme by Stokes and Einstein *J. Chem. Phys.* 138 12A526

[29] Kumar P, Buldyrev S V, Becker S R, Poole P H, Starr F W and Stanley H E 2007 Relation between the Widom line and the breakdown of the Stokes–Einstein relation in supercooled water *Proc. Natl. Acad. Sci. U. S. A.* 104 9575

[30] Lewis L J and Wahnström G 1994 Molecular-dynamics study of supercooled ortho-terphenyl *Phys. Rev. E* 50 3865

[31] Binder K and Kob W 2011 *Glassy materials and disordered solids: An introduction to their statistical mechanics* World Scientific)

[32] Mazza M G, Giovambattista N, Starr F W and Stanley H E 2006 Relation between Rotational and Translational Dynamic Heterogeneities in Water *Phys. Rev. Lett.* 96 057803

[33] Kämmerer S, Kob W and Schilling R 1997 Dynamics of the rotational degrees of freedom in a supercooled liquid of diatomic molecules *Phys. Rev. E* 56 5450

[34] Mishra C K and Ganapathy R 2015 Shape of Dynamical Heterogeneities and Fractional Stokes-Einstein and Stokes-Einstein-Debye Relations in Quasi-Two-Dimensional Suspensions of Colloidal Ellipsoids *Phys. Rev. Lett.* 114 198302